\documentclass[11pt]{article}
\newcommand{\be}{\begin{equation}}
\newcommand{\ee}{\end{equation}}
\newcommand{\bea}{\begin{eqnarray}}
\newcommand{\eea}{\end{eqnarray}}

\def\E{\> = \>}

\begin{document}


\thispagestyle{empty}
\setcounter{page}{0}

\vspace*{2cm}

\begin{center}
{\large\bf Coulomb Corrections to Elastic Electron-Proton Scattering and the 
Proton Charge Radius}

\vspace{1.5cm}

\renewcommand{\thefootnote}{\fnsymbol{footnote}}

R.~Rosenfelder \footnote[1]{E-mail address: rosenfelder@psi.ch}

\renewcommand{\thefootnote}{\alph{footnote}}

\vspace{1cm}

Paul Scherrer Institut, CH-5232 Villigen PSI, Switzerland

\end{center}

\vspace{5cm}
\begin{abstract}
\noindent
Recent values of the proton charge radius 
derived from the Lamb shift in electronic hydrogen  
tend to be larger than those from electron scattering. Therefore
low-momentum-transfer scattering data from different groups and laboratories 
have been reanalyzed with Coulomb and recoil corrections included. This 
was done by calculating at each scattering angle and scattering energy 
the corresponding correction 
obtained from a partial wave program which includes recoil approximately.
Corrections due to magnetic scattering were also made
before the rms-radius was determined by a fit which allows free 
normalization for each experiment.
It is shown that an analysis of electron scattering data 
including Coulomb corrections lowers the $\chi^2$ of the fit and 
increases the proton radius 
by about $ (0.008 - 0.013) $ fm depending on the fit strategy. 
A value of $ r_p = 0.880\, (15) $ fm is 
obtained which is in good agreement with the one extracted from Lamb shift 
measurements.

\end{abstract}

\newpage

\setcounter{equation}{0}

\noindent
{\bf 1.} 
Traditionally elastic electron scattering data from the proton have been 
analyzed in the first Born (one-photon-exchange) approximation since 
$Z \alpha \simeq 1/137$  
was supposed to be small enough. In this way a root-mean-square charge radius of
\be
 <r^2>^{1/2}_{\rm e-scatt} \E 0.862\, (12) \> \> \> {\rm fm}
\label{Mainz proton radius}
\ee
has been deduced from elastic electron-proton scattering experiments at Mainz 
\cite{Simon1} 20 years ago. This value superseded the nearly canonical 
value of $ 0.805 (11)$ fm from Stanford \cite{Hand} which was based on an (invalid) 
$Q^2 \to 0$ extrapolation of the dipole form factor fitted to data at higher 
momentum transfer~\footnote{Although the deviation of the data at low momentum 
transfer from the standard dipole parametrization is clearly visible and unambigous 
(see e.g. Fig. 4 in ref. \cite{Bork2}) the Stanford value is still cited as an 
alternative in the 
(atomic and high energy physics) literature. However, the precise absolute e-p 
cross sections measured by the Walther group at Mainz
have been confirmed subsequently in other measurements and laboratories
(most lately at Saclay, see Fig. 2 in ref. \cite{Saclay} ) so that there is 
no convincing reason for doing so.}. In the last few years
atomic Lamb shift measurements in hydrogen have become increasingly more accurate 
and are now very sensitive to the proton size \cite{Lamb}.
Assuming the validity of bound state QED calculations a value for the proton charge
radius can be deduced from them; the most recent theoretical evaluation \cite{QED} 
leads to
\be
 <r^2>^{1/2}_{\rm Lamb \> shift} \E 0.883 \, (14) \> \> \> {\rm fm} 
\label{Lamb shift proton radius}
\ee
which is slighthly higher than the value (\ref{Mainz proton radius}).
A possible reason of this discrepancy may be the distortion of the scattering
waves in the electrostatic field of the proton. Indeed, Coulomb corrections 
in elastic electron scattering have been shown recently to {\it increase} the
rms charge radius of the $Z = 1, N = 1$ hydrogen isotope, the deuteron by 
$0.012 - 0.015$  fm \cite{deut1},\cite{deut2}. Therefore it is of some interest 
to investigate whether Coulomb corrections also affect the determination of the 
proton radius. Eventually  a 
muonic hydrogen experiment which has started at PSI \cite{PSI-R-98-03} will 
yield this fundamental hadronic quantity with a tenfold higher accuracy  
than given in (\ref{Mainz proton radius}) or in (\ref{Lamb shift proton radius}). 
Even then it will be useful to see whether the value from electron scattering is 
consistent with the precise value.

\vspace{0.3cm}
\noindent
{\bf 2.}
In the one-photon-exchange approximation the elastic electron-proton scattering 
cross section in the lab frame is given by
\be
\frac{d \sigma}{d \Omega} \Bigg |_{\rm 1-photon} \E \sigma_{\rm Mott} \, 
f_{\rm recoil} \, \left [  \>
\frac{G_E^2(Q^2) + \tau G_M^2(Q^2)}{1+\tau} + 2 \tau G_M^2(Q^2) \tan^2 \left (
\frac{\Theta}{2} \right ) \, \right ] 
\label{one photon}
\ee
where $ \tau = Q^2/4 M_p, Q^2 = 4 E_i E_f \sin^2(\Theta/2) $. $E_i, E_f, \Theta$ are 
the initial, final electron energy and the scattering angle, respectively.
The recoil factor is given by $ 
f_{\rm recoil} = \left [ 1 + 2 E_i \sin^2 ( \Theta/2 )/M_p  \right ]^{-1} $
and $G_{E/M}(Q^2)$ denote the electric and magnetic (Sachs) form factors. The former
has the low $Q^2$ expansion
\be
G_E(Q^2) \E 1 - \frac{1}{6} \left < r^2 \right > Q^2 + \frac{1}{120} 
\left < r^4 \right > Q^4 + \ldots
\ee
from which the root-mean-square charge radius 
$ r_p \equiv \left < r^2 \right >^{1/2}  $ of the 
proton can be determined. At low momentum transfer the magnetic form factor is only 
a small correction and and a common approximation is to assume 
$G_M(Q^2) = \mu_p \, G_E(Q^2) $ where $\mu_p = 2.793 $ is
the magnetic moment of the proton. Therefore the full cross section may be written 
as 
\be
\frac{d \sigma}{d \Omega} \E \sigma_{\rm Mott} \, f_{\rm recoil} \, 
\frac{G_E^2(Q^2)}{1+\tau} \left (  \> 1 + \Delta_{\rm mag} \> \right ) \, \left ( \> 
1  + \Delta_c \> \right )
\label{with Coul}
\ee
where 
\be
 \Delta_{\rm mag} \E \frac{G_M^2(Q^2)}{G_E^2(Q^2)} \, \tau \left [ \, 1 + 2 \, (1 + \tau ) 
\tan^2 \left (\frac{\Theta}{2} \right ) \,  \right ]  \> .
\ee
Here it is assumed that the Coulomb corrections affect the magnetic part in the 
same way as the electric part; in any case eq. (\ref{with Coul}) is correct if 
both corrections are small.

To evaluate the Coulomb correction $ \Delta_c $ one may either calculate the
second Born approximation or use a standard partial wave program which
solves the Dirac equation in the electrostatic potential $V_c$ of the proton. 
Here I adopt the second approach which is numerically easier since it does not 
require principal value integrations. The drawback is that magnetic scattering 
is not included and that it is difficult to implement 
recoil corrections properly which naturally are more important for
electron scattering from the proton than from heavier targets. Following 
ref. \cite{Fri} I transform to the center-of-mass system but neglect dynamical 
recoil corrections, i.e. the
additional interaction potential $\Delta V$ which is of order $V_c^2/M_p$. 
This is a standard procedure since the work of Foldy {\it et al.} \cite{FFY} and
has been shown to produce Coulomb corrections at low momentum transfer which 
are very close to the one from evaluating the real part of the second-order Born 
amplitude \cite{deut2}.
Of course, the Coulomb corrections require knowledge of $V_c(r)$, i.e. of 
the whole charge distribution and not just of its second moment. From 
eq. (\ref{with Coul}) one can read off
\be
\rho_c(r) \E \int \frac{d^3q}{(2 \pi)^3} \> \frac{G_E({\bf q}^2)}{\sqrt{1 + 
{\bf q}^2/(4 M_p^2)}} \, e^{i {\bf q} \cdot {\bf r}} \> ; \hspace{0.3cm}
V_c(r) \E \int d^3x \> \frac{\rho_c(x)}{|{\bf r} - {\bf x} |}
\label{def rho}
\ee
if the leading term is taken as square of an elastic form factor which itself is 
the Fourier transform of a charge distribution. Since (in the Breit frame) the Sachs 
form factors can indeed be interpreted as arising from appropriate distributions, 
Eq. (\ref{def rho}) just describes the folding of this distribution with the
one from the Darwin-Foldy term. For $G_E({\bf q}^2)$~~I have used the parametrization 
given in eq. (9) and Table 3 of ref. \cite{Simon1} which describes the form factor
data over a wide range of $Q^2$ and which corresponds to a sum of Yukawa
charge distributions. 
The folding was performed by a Fourier-Bessel
expansion of the distribution with a cut-off radius of $5$ fm and 
up to 300 terms. The Coulomb correction for each scattering angle and energy
is then given by
\be
\Delta_c^{(i)} \E \frac{\sigma_{\rm partial \> wave}(E_i,\Theta_i)}
{\sigma_{\rm charge, \> Born}(E_i,\Theta_i)} - 1 
\ee
where $\sigma_{\rm charge, \> Born}$ is the cross section in Born approximation 
due to the charge distribution $\rho_c(r)$.
For small scattering angles the Coulomb corrections are all 
{\it positive} leading to an increase of the rms radius \cite{deut2}. 

\vspace{0.3cm}
\noindent
{\bf 3.} I have (re)analyzed the old e-p cross section data 
below $Q^2 < 1$ fm$^{-2}$ from Mainz and Saskatoon. More recent experiments
are either at higher $Q^2$ \cite{SLAC} or measure form factor ratios by polarization 
transfer \cite{ratio}.
The Mainz data include published cross sections from the 
Walther group \cite{Simon1,Simon2,Bork1,Bork2} as well as unpublished data 
contained in the PhD thesis by W. Reuter \cite{Reut}. The Saskatoon experiment 
\cite{Sask} detected the struck proton but the data can be easily converted to 
electron angles. In the following these experimental data sets will be denoted as 
S1, S2, B1, B2, R and M, respectively.
For ease of comparison and further reference the cross sections, their 
quoted errors, and the corrections
$\Delta_{\rm mag}, \Delta_c $ are collected in Table 1 as well as the resulting 
values of the electric form factor with errors. Note that in the latest two 
experiments by Simon {\it et al.} (abbreviated S1, S2) the systematic errors 
were added linearly
to the random errors whereas the other experiments do not specify how their 
systematic errors are treated. 
In view of that their total cross section errors may be underestimated.
The correction due to magnetic scattering has been obtained with the parametrizations 
for $G_{M/E}$ as given in ref. \cite{Simon1}; its relative error was
estimated to be  $3\cdot 10^{-3}$ by comparing with the ansatz $ G_M = G_E$.
From Table 1 one sees that the Coulomb corrections 
are small but positive, ranging from $0.4$ -- $0.9 \%$ which is roughly the size of the 
statistical or systematic errors in the experiments of Simon {\it et al.}
By comparing with the result of a phase shift calculation, which uses an exponential 
distribution with the same rms-radius, an error of at most $2 \cdot 10^{-4}$ can be 
ascribed to the Coulomb corrections listed in Table 1. 
Qualitatively one therefore expects that the analysis leads to an upward shift of the 
proton radius of roughly the order of the error which was quoted in previous analyses.

This is indeed born out by a fit of the electric form factor 
at low momentum transfers to the expression
\be
G_E^{\> \rm ansatz} (Q^2) \E a_0  + a_1 Q^2 + a_2 Q^4 \> .
\label{ansatz GE}
\ee
It has been recognized very early that the normalization errors $\delta_{\rm norm}$  in 
absolute cross section measurements require a {\it free} normalization factor $a_0$ 
in eq. (\ref{ansatz GE}). Wong \cite{Won} has further emphasized 
that the normalization errors have to be part of the $\chi^2$-function
\be
\chi^2 \E \sum_{i=1}^n \left ( \frac{G_E^{(i) \> \> {\rm exp}} - 
G_E^{(i) \> \> {\rm ansatz}}}{\delta G_E^{(i) \> \> {\rm exp}}} \right )^2 +  
\sum_{j} \left ( \frac{1 - a_0^{(j)}}{\frac{1}{2} \delta_{\rm norm}^{(j)}} \right )^2
\label{chi2}
\ee
where $i$ runs over the $n$ experimental points and 
$ j $ over the different experiments. The factor $1/2$ in the 
denominator of the last term in eq. (\ref{chi2}) enters because the form factor and 
not the cross section  is fitted. According to Wong's 
analysis of the Mainz data, ``the result would have been 
\be
r_p^{\rm Wong} \E 0.877 \, (24) \> \> {\rm fm}
\label{Wong proton radius}
\ee
with $a_0 = 1.0028 (22) $ ''. 
Since only low $Q^2$-data are analyzed the coefficient $a_2$ is not determined
well by the fit. A value $ a_2/a_0 = 0.011 \pm 0.004 \> $ fm$^4$
has been determined in ref. \cite{Bork3}. However, since the S1-parametrization of $G_E$
is a very good representation of the available
e-p data \footnote{Platchkov {\it et al.}  \cite{Saclay} have found that their
absolute proton cross sections measured in 
the momentum transfer region between 1 and 16 fm$^{-2}$ ``agree within $1 \%$ with the 
four-pole proton form factor parametrization of Simon {\it et al.}''} 
\, I rather adopt its value of $a_2 = 0.0141 \>  $ fm$^4$ and use in most of the fits
\be
\frac{a_2}{a_0} \E \left ( 0.014 \pm 0.004 \right ) \> \> \> {\rm fm}^4  \> \> .
\label{value a2}
\ee
Note that the error assumed above corresponds to
a $ 7 \% $  error in the moment $ <r^4>^{1/4} $ which probably is a generous 
estimate. This error (muliplied by $Q^4$) is now added linearly to the errors
of $G_E$ listed in Table 1.

\vspace{0.3cm}
\noindent
{\bf 4.} The results of fitting  eq. (\ref{ansatz GE}) to the data collected in Table 1 
are given in Table 2, both without (case A) and with (case B) Coulomb corrections.
Several fits have been made differing in additional constraints for the size of the
corrections. One constraint already used in Table 1 is a low momentum tranfer so that
the recoil corrections do not become too large; another one is the dominance 
of charge scattering for which the Coulomb corrections are most reliable.
In nearly all cases the magnetic contributions were therefore restricted to be less 
than $10 \%$. 
Fit 3 gives the results when this is relaxed to $\Delta_{\rm mag} < 15 \% $.
The result of fit 1\,A is very close to the original Mainz proton radius  
(\ref{Mainz proton radius}) although it uses an unrealistic small value of $a_2$.
In fit 4 only the data of the Walther group are analyzed; without Coulomb 
corrections this indeed gives a value of the proton radius which is in agreement with 
Wong's one in eq. (\ref{Wong proton radius}). However, for unknown reasons 
I do not observe that ``uncertainties in the normalization factor ... have significantly 
increased the error in the deduced radius''\footnote{The errors given in Table 2 are the
``naive'' errors obtained from the MINUIT fit program.}. 
Fit 5 omits the more indirect Sakatoon data (abbreviated by M) which
give the largest individual contribution to the $\chi^2$. Finally in fit 6 the errors
in data sets R, B1, B2, M were artificially enlarged by a factor of 2 to take into 
account systematic errors which are not included in the original data.
 
In all cases one observes that
the inclusion of Coulomb corrections {\it improves} the $\chi^2$ of the fit and 
{\it increases} the deduced proton radius by 
\be
\Delta r_p \E ( 0.008 - 0.013 ) \> \> {\rm fm }  
\label{delta rp from Coul}
\ee
depending on the different fit strategies adopted. For the fits with 
$\Delta_{\rm mag} < 0.1 $ the range is narrower: 
$\Delta r_p = ( 0.010 - 0.013 ) $ fm. Fits 5\,B and 6\,B may be considered the
most realistic ones and give
\be
< r^2 >^{1/2}_{\rm e-scatt \> with \> Coulomb} \E 
\left ( 0.880 \pm 0.015 \right ) \> \> {\rm fm} \> .
\label{proton radius with Coul}
\ee
Here I have slightly enlarged the errors obtained in both fits 
\footnote{Historically there seems to be a tendency to underestimate the (systematic)
errors in the proton radius determined from electron scattering and the above
value may also be insufficient in this respect. However, a thorough treatment of 
systematic errors is beyond the scope of the present investigation which focuses
on the change of the extracted radius due to Coulomb corrections.}. 
Eq. (\ref{proton radius with Coul}) is now in good 
agreement with the value (\ref{Lamb shift proton radius}) deduced from the recent
Lamb shift measurements.

\vspace{8.5cm}

\noindent
{\bf Acknowledgements :} I remember with gratitude and some sadness many 
discussions with the late V. Walther and members of his group at Mainz University. 
His untimely death cut short a very productive and farseeing research.
At PSI I would like to thank David Taqqu for a critical reading of the manuscript.

\newpage

\thispagestyle{empty}
\begin{table}[p]
\begin{center}
\caption {Electron-proton scattering experiment, incident energy, scattering 
angle, four momentum transfer squared and 
measured cross section with error (in units of the last digit). $\delta_{\rm norm}$
denotes the normalization error as given in the references.
$\Delta_{\rm mag}$ is the correction for magnetic scattering, $\Delta_c$
the Coulomb correction (see text). The final column gives the electric form 
factor and its error, again in units of the last digit. Entries which are blank
have the same value as in the previous line.}

\begin{tabular}{|l|l|l|c|r|c|c|c|c|} \hline
     &                &                             &                            &          
                                                    &                            &
                      &                             &                                  \\ 
Exp. & $E_i\> $ [MeV] & $\> \> \>\Theta \> [^{\circ}]$ & $Q^2 \> $ [fm$^{-2}$]   & 
$\sigma \> \> (\delta \sigma)$ \hspace{0.2cm}       &   $\delta_{\rm norm}$      &
$\Delta_{\rm mag}$ & $\Delta_c         $            &   $ G_E \> \> (\delta G_E)$         \\ 
     &             &                                &                            &       
                   $[10^{-4}$ fm$^2$/sr]            &                            &
                   &                                &                                  \\ 
\hline
   &        &        &        &               &            &         &         &             \\
S1 & 149.84 & 27.965 & 0.1322 &  60.72  (106) & 0.0048$^a$ & 0.0128 & 0.0045 & 0.9770 (86) \\
   & 228.65 &        & 0.3049 &  25.47~ ~(27) &            & 0.0296 & 0.0042 & 0.9634 (52) \\
   & 274.83 &        & 0.4380 &  17.25~ ~(18) &            & 0.0426 & 0.0040 & 0.9504 (51) \\
   & 301.42 &        & 0.5252 &  13.95~ ~(13) &            & 0.0511 & 0.0039 & 0.9356 (45) \\
   & 321.26 &        & 0.5952 &  12.10~ ~(10) &            & 0.0580 & 0.0038 & 0.9272 (40) \\
S2 & 149.4  & 28     & 0.1317 &  61.96~ ~(72) & 0.0056$^b$ & 0.0128 & 0.0045 & 0.9866 (58) \\
   &        & 30     & 0.1504 &  46.19~ ~(47) &            & 0.0149 & 0.0047 & 0.9797 (51) \\
   &        & 35     & 0.2015 &  24.51~ ~(23) &            & 0.0209 & 0.0052 & 0.9763 (47) \\
   &        & 40     & 0.2586 &  13.94~ ~(14) &            & 0.0283 & 0.0055 & 0.9674 (50) \\
   &        & 45     & 0.3208 &   8.43~ ~~(8) &            & 0.0373 & 0.0058 & 0.9582 (47) \\
   &        & 50     & 0.3875 &   5.403 ~(52) &            & 0.0481 & 0.0061 & 0.9536 (47) \\
   &        & 55     & 0.4578 &   3.598 ~(40) &            & 0.0612 & 0.0063 & 0.9485 (54) \\
   &        & 60     & 0.5310 &   2.428 ~(25) &            & 0.0768 & 0.0064 & 0.9344 (50) \\
   &        & 65     & 0.6062 &   1.718 ~(18) &            & 0.0954 & 0.0066 & 0.9296 (51) \\
   &        & 70     & 0.6828 &   1.225 ~(14) &            & 0.1176 & 0.0066 & 0.9175 (55) \\
   & 199.5  & 28     & 0.2335 &  33.64~ ~(35) &            & 0.0227 & 0.0044 & 0.9695 (52) \\
   &        & 30     & 0.2663 &  25.31~ ~(26) &            & 0.0263 & 0.0045 & 0.9670 (51) \\
   &        & 35     & 0.3560 &  13.34~ ~(14) &            & 0.0369 & 0.0049 & 0.9598 (52) \\
   &        & 40     & 0.4556 &   7.478 ~(66) &            & 0.0499 & 0.0051 & 0.9432 (43) \\
   &        & 45     & 0.5637 &   4.493 ~(41) &            & 0.0656 & 0.0053 & 0.9300 (44) \\
   &        & 50     & 0.6787 &   2.865 ~(27) &            & 0.0845 & 0.0054 & 0.9216 (45) \\
   &        & 55     & 0.7993 &   1.866 ~(22) &            & 0.1072 & 0.0055 & 0.9045 (56) \\
   &        & 60     & 0.9239 &   1.283 ~(17) &            & 0.1341 & 0.0055 & 0.8970 (62) \\
R  & 100    & 30     & 0.0678 &104.9~~ (10.5) &0.011$^c$  & 0.0067 & 0.0048 & 0.9883 (51) \\
   &        & 40     & 0.1172 &  31.71~ ~(25) &            & 0.0128 & 0.0058 & 0.9772 (40) \\
   &        & 50     & 0.1767 &  12.54~ ~(10) &            & 0.0219 & 0.0066 & 0.9747 (40) \\
   &        & 60     & 0.2438 &   5.805 ~(70) &            & 0.0351 & 0.0072 & 0.9723 (60) \\
   &        &        &        &   5.751 ~(46) &            &        &        & 0.9677 (40) \\
   &        &        &        &   5.768 ~(69) &            &        &        & 0.9692 (59) \\
   &        & 70     & 0.3158 &   2.934 ~(29) &            & 0.0542 & 0.0076 & 0.9607 (49) \\
   &        & 80     & 0.3901 &   1.627 ~(16) &            & 0.0814 & 0.0080 & 0.9567 (49) \\
   &        & 90     & 0.4642 &   0.9478(76)  &            & 0.1209 & 0.0082 & 0.9486 (41) \\
   & 300    & 28     & 0.5216 &  14.01~ ~(14) &            & 0.0508 & 0.0039 & 0.9356 (48) \\
\hline
\end{tabular}
\end{center}
\end{table}
\newpage

\begin{table}[p]
\begin{center}
Table 1 (continued)

\vspace{0.2cm} 

\begin{tabular}{|l|l|l|c|r|c|c|c|c|} \hline
     &                &                             &                            &          
                                                    &                            &
                      &                             &                                  \\ 
Exp. & $E_i\> $ [MeV] & $\> \> \>\Theta \> [^{\circ}]$ & $Q^2 \> $ [fm$^{-2}$]   & 
$\sigma \> \> (\delta \sigma)$  \hspace{0.2cm}      &   $\delta_{\rm norm}$      &
$\Delta_{\rm mag}$ & $\Delta_c         $            &   $ G_E \> \> (\delta G_E)$         \\ 
     &             &                                &                            &       
                   $[10^{-4}$ fm$^2$/sr]            &                            &
                   &                                &                                  \\ 
\hline
   &        &        &        &              &            &        &        &             \\
B1 & 149.8  & 28     & 0.1324 &  60.9~~ (12) & 0.018$^d$  & 0.0129 & 0.0045 & 0.9807 (98) \\
   &        & 40     & 0.2599 &  13.81~ (13) &            & 0.0284 & 0.0055 & 0.9654 (47) \\
   &        & 45     & 0.3225 &   8.22~ ~(8) &            & 0.0375 & 0.0058 & 0.9487 (48) \\
   &        & 50     & 0.3895 &   5.27~ ~(6) &            & 0.0484 & 0.0061 & 0.9443 (55) \\
   &        & 55     & 0.4602 &   3.49~ ~(3) &            & 0.0615 & 0.0063 & 0.9366 (42) \\
   &        & 60     & 0.5337 &   2.40~ ~(2) &            & 0.0772 & 0.0064 & 0.9314 (41) \\
   &        & 65     & 0.6093 &   1.668 (14) &            & 0.0959 & 0.0066 & 0.9183 (41) \\
   &        & 70     & 0.6863 &   1.197 (12) &            & 0.1182 & 0.0066 & 0.9093 (48) \\
   &        & 75     & 0.7639 &   0.871 (10) &            & 0.1448 & 0.0067 & 0.8973 (54) \\
   & 180.1  & 28     & 0.1907 &  41.8~~ ~(8) &            & 0.0185 & 0.0044 & 0.9763 (95) \\
   &        & 40     & 0.3730 &   9.28~ ~(8) &            & 0.0408 & 0.0053 & 0.9499 (42) \\
   &        & 45     & 0.4620 &   5.55~ ~(5) &            & 0.0538 & 0.0055 & 0.9350 (44) \\
   &        & 50     & 0.5569 &   3.56~ ~(3) &            & 0.0693 & 0.0057 & 0.9300 (41) \\
   &        & 55     & 0.6567 &   2.316 (17) &            & 0.0879 & 0.0058 & 0.9132 (36) \\
   &        & 60     & 0.7601 &   1.588 (11) &            & 0.1101 & 0.0059 & 0.9053 (34) \\
   &        & 65     & 0.8660 &   1.107 ~(8) &            & 0.1366 & 0.0059 & 0.8922 (35) \\
   &        & 70     & 0.9733 &   0.799 ~(7) &            & 0.1681 & 0.0059 & 0.8839 (42) \\
   & 199.6  & 28     & 0.2337 &  33.2~~ ~(7) &            & 0.0227 & 0.0044 & 0.9637 (103)\\
   &        & 40     & 0.4561 &   7.32~ (10) &            & 0.0500 & 0.0051 & 0.9336 (65) \\
   &        & 45     & 0.5642 &   4.44~ ~(5) &            & 0.0657 & 0.0053 & 0.9249 (54) \\
   &        & 50     & 0.6794 &   2.78~ ~(4) &            & 0.0846 & 0.0054 & 0.9082 (67) \\
   &        & 60     & 0.9248 &   1.247 (14) &            & 0.1342 & 0.0055 & 0.8847 (52) \\
   & 228.9  & 28     & 0.3063 &  24.8~~ ~(5) &            & 0.0298 & 0.0042 & 0.9540 (98) \\
   &        & 40     & 0.5956 &   5.51~ ~(7) &            & 0.0653 & 0.0048 & 0.9262 (61) \\
   &        & 45     & 0.7357 &   3.26~ ~(4) &            & 0.0858 & 0.0050 & 0.9053 (58) \\
   &        & 50     & 0.8843 &   2.06~ ~(2) &            & 0.1103 & 0.0050 & 0.8919 (46) \\
   & 249.6  & 28     & 0.3633 &  20.2~~ ~(4) &            & 0.0353 & 0.0042 & 0.9378 (94) \\
   &        & 35     & 0.5521 &   7.89~ ~(9) &            & 0.0574 & 0.0045 & 0.9200 (54) \\
   &        & 40     & 0.7048 &   4.46~ ~(6) &            & 0.0774 & 0.0046 & 0.9064 (63) \\
   &        & 45     & 0.8695 &   2.66~ ~(3) &            & 0.1015 & 0.0047 & 0.8888 (52) \\
   & 275.3  & 28     & 0.4405 &  16.84~ (34) &            & 0.0429 & 0.0040 & 0.9430 (97) \\
\hline
\end{tabular}
\end{center}
\end{table}
\newpage

\begin{table}[p]
\begin{center}
 Table 1 (continued)

\vspace{0.2cm}

\begin{tabular}{|l|l|l|c|r|c|c|c|c|} \hline
     &                &                             &                            &          
                                                    &                            &
                      &                             &                                  \\ 
Exp. & $E_i\> $ [MeV] & $\> \> \>\Theta \> [^{\circ}]$ & $Q^2 \> $ [fm$^{-2}$]   & 
$\sigma \> \> (\delta \sigma)$   \hspace{0.2cm}     &   $\delta_{\rm norm}$      &
$\Delta_{\rm mag}$ & $\Delta_c         $            &   $ G_E \> \> (\delta G_E)$         \\ 
     &             &                                &                            &       
                   $[10^{-4}$ fm$^2$/sr]            &                            &
                   &                                &                                  \\ 
\hline
   &        &        &        &              &            &        &        &             \\ 
B2 & 149.8  & 80     & 0.8414 &   0.659 ~(6) & 0.018$^d$  & 0.1766 & 0.0067 & 0.8947 (44) \\
   &        & 90     & 0.9939 &   0.386 ~(3) &            & 0.2606 & 0.0067 & 0.8786 (38) \\
   & 228.9  & 28     & 0.3063 &  24.5~~ ~(5) &            & 0.0298 & 0.0042 & 0.9482 (98) \\
   &        & 40     & 0.5956 &   5.46~ ~(7) &            & 0.0653 & 0.0048 & 0.9220 (61) \\
   &        & 45     & 0.7357 &   3.23~ ~(4) &            & 0.0858 & 0.0050 & 0.9011 (58) \\
   &        & 50     & 0.8843 &   2.04~ ~(2) &            & 0.1103 & 0.0050 & 0.8876 (46) \\
   & 249.6  & 28     & 0.3633 &  20.4~~ ~(4) &            & 0.0353 & 0.0042 & 0.9425 (94) \\
   &        & 35     & 0.5521 &   7.97~ ~(9) &            & 0.0574 & 0.0045 & 0.9246 (54) \\
   &        & 40     & 0.7048 &   4.51~ ~(6) &            & 0.0774 & 0.0046 & 0.9115 (63) \\
   &        & 45     & 0.8695 &   2.69~ ~(3) &            & 0.1015 & 0.0047 & 0.8938 (52) \\
   & 275.3  & 35     & 0.6685 &   6.43~ ~(7) &            & 0.0695 & 0.0043 & 0.9136 (52) \\
   &        & 40     & 0.8523 &   3.59~ ~(4) &            & 0.0936 & 0.0044 & 0.8937 (52) \\
   & 298.5  & 28     & 0.5165 &  13.92~ (15) &            & 0.0503 & 0.0039 & 0.9280 (52) \\
   &        & 30     & 0.5881 &  10.34~ (11) &            & 0.0583 & 0.0040 & 0.9188 (51) \\
   &        & 35     & 0.7826 &   5.31~ ~(6) &            & 0.0814 & 0.0041 & 0.8978 (53) \\
   &        & 40     & 0.9966 &   2.96~ ~(3) &            & 0.1096 & 0.0041 & 0.8768 (47) \\
M  &  57.3  & 86.61  & 0.1500 &   3.557 (36) & 0.046 $^e$ & 0.0359 & 0.0090 & 0.9768 (51) \\
   &  82.2  & 85.19  & 0.2943 &   1.833 (19) &            & 0.0683 & 0.0085 & 0.9647 (52) \\
   &  82.4  & 85.18  & 0.2957 &   1.815 (11) &            & 0.0686 & 0.0085 & 0.9620 (31) \\
   &  89.7  & 84.78  & 0.3456 &   1.555 (10) &            & 0.0795 & 0.0084 & 0.9571 (33) \\
   &  95.6  & 84.45  & 0.3883 &   1.401 ~(9) &            & 0.0887 & 0.0082 & 0.9582 (33) \\
   &  96.4  & 84.41  & 0.3943 &   1.362 ~(9) &            & 0.0900 & 0.0082 & 0.9514 (34) \\
   & 102.2  & 84.09  & 0.4384 &   1.227 ~(8) &            & 0.0994 & 0.0080 & 0.9474 (33) \\
   & 108.7  & 83.73  & 0.4900 &   1.089 ~(9) &            & 0.1103 & 0.0079 & 0.9381 (41) \\
   &  90.4  &115.33  & 0.5269 &   0.349 ~(7) &            & 0.2730 & 0.0089 & 0.9345 (98) \\
   & 118.6  & 93.23  & 0.6733 &   0.552 ~(5) &            & 0.1887 & 0.0077 & 0.9189 (45) \\
   & 129.5  & 92.64  & 0.7874 &   0.475 ~(4) &            & 0.2177 & 0.0074 & 0.9107 (42) \\
   &        &        &        &              &            &        &        &             \\ 
\hline
\end{tabular}
\end{center}
\vspace{0.1cm}
 
S1: ref. \cite{Simon1}, Table 1, systematic errors added linearly to statistical 
cross section errors \\
S2: ref.  \cite{Simon2}, Table 1,
systematic error of $0.46 \%$ added linearly to statistical cross section errors \\
R: ref. \cite{Reut}, Tabelle 11, cross section errors from $G_E$ errors \\
B1: ref. \cite{Bork1}, Table 1 ;
\hspace{0.3cm} B2: ref. \cite{Bork2}, Table 1 \\
M: ref. \cite{Sask}, Table II, cross sections and errors from 
$G_E, \delta G_E $ values 
\vspace{0.1cm}

$^a$ caption to Table 1 in S1;\hspace{0.1cm} $^b$ caption to Table 1 in S2;
\hspace{0.1cm} $^c$ p. 92, line 14 in R;
\hspace{0.1cm} $^d$  p. 387, line 4 in S1;\\
$^e$ Table I in M 
\vspace{0.1cm}

$\alpha^{-1} = 137.036, \> \hbar c = 197.327 $ MeV fm, $ \> M_p = 938.272$ MeV,
$ \> \mu_p = 2.79285 $ .

\end{table}

\begin{table}[htb]
\begin{center}
Table 2: Fit results: A without, B with Coulomb corrections. Errors are in units of the 
last digit. 
\vspace{0.1cm}

\begin{tabular}{|c|c|c|c|} \hline
                   &                       &                       &                     \\ 
Fit                &         1             &           2           &          3          \\
                   &  A \quad\quad\quad B  &  A \quad\quad\quad B  & A \quad\quad\quad B  \\
                   &                       &                       &                      \\ 
\hline
                   &                       &                       &                     \\
$\Delta_{\rm mag}$ &       $<$ 0.1         &        $<$ 0.1        &         $<$ 0.15     \\
$n$                &         70            &           70          &             85       \\
$a_2$ [fm$^4$]     &       0.011(4)        &         0.014 (4)      &           0.014 (4)   \\
$\Delta_c$         & 0 \hspace{1.1cm} $\neq 0$ & 0 \hspace{1.1cm} $\neq 0$ 
                   & 0 \hspace{1.1cm} $\neq 0$ \\
$\chi^2$           &  22.00 ~\quad~  20.97 &  22.39 ~\quad~ 21.08  & 26.83~ \quad~ 24.68   \\
$a_0$(S1)          & 1.0002(19)~~0.9999(20)& 1.0003(19)~~1.0000(20)& 1.0004(19)~~1.0000(19)\\
$a_0$(S2)          & 1.0014(17)~~1.0002(17)& 1.0016(17)~~1.0004(17)& 1.0017(16)~~1.0004(16)\\
$a_0$(R)~          & 1.0002(18)~~0.9979(18)& 1.0004(18)~~0.9981(18)& 1.0013(17)~~0.9988(17)\\
$a_0$(B1)          & 0.9938(22)~~0.9926(22)& 0.9940(22)~~0.9928(22)& 0.9939(21)~~0.9926(21)\\
$a_0$(B2)          & 0.9910(32)~~0.9908(32)& 0.9910(32)~~0.9908(32)& 0.9914(29)~~0.9909(30)\\
$a_0$(M)           & 1.0033(20)~~1.0002(21)& 1.0036(20)~~1.0005(21)& 1.0036(19)~~1.0003(19)\\
                   &                       &                       &                       \\
$r_p$ [fm]     & 0.861 (12)~~~0.871 (12) & 0.867 (12)~~~0.878 (12) & 0.870 (10)~~~0.878 (10) \\
                   &                       &                       &                       \\
\hline  
\end{tabular}
\vspace{0.4cm}

Table 2 (continued)
\vspace{0.1cm}

\begin{tabular}{|c|c|c|c|} \hline
                   &                       &                       &                      \\ 
Fit                &           4           &           5           &           6          \\
                   &  A \quad\quad\quad B  &  A \quad\quad\quad B  &  A \quad\quad\quad B  \\
                   &                       &                       &                       \\
\hline
                   &                       &                       &                      \\
$\Delta_{\rm mag}$ &        $<$ 0.1        &         $<$ 0.1       &         $<$ 0.1      \\
$n$                &       54$^a$          &         63$^b$        &            70        \\
$a_2$ [fm$^4$]     &      0.014 (4)         &        0.014 (4)       &          0.014 (4)    \\
$\Delta_c$         & 0 \hspace{1.1cm} $\neq 0$ & 0 \hspace{1.1cm} $\neq 0$ 
                   & 0 \hspace{1.1cm} $\neq 0$ \\
$\chi^2$           &  14.29 ~\quad ~13.03  &  19.52 ~\quad~17.88 & ~9.96$^c$~\quad~~9.29$^c$\\
$a_0$(S1)          & 1.0007(20)~~1.0004(20)& 1.0004(19)~~1.0001(20)& 1.0003(20)~~1.0002(20)\\
$a_0$(S2)          & 1.0022(17)~~1.0010(17)& 1.0018(17)~~1.0006(17)& 1.0016(17)~~1.0006(18)\\
$a_0$(R)~          &           -           & 1.0006(18)~~0.9983(18)& 1.0004(29)~~0.9988(29)\\
$a_0$(B1)          & 0.9950(23)~~0.9938(23)& 0.9943(22)~~0.9932(22)& 0.9944(31)~~0.9936(32)\\
$a_0$(B2)          & 0.9922(33)~~0.9921(33)& 0.9914(32)~~0.9913(32)& 0.9923(45)~~0.9926(45)\\
$a_0$(M)           &    -                  &         -             & 1.0037(33)~~1.0009(34)\\
                   &                       &                       &                      \\
$r_p$ [fm]     & 0.875 (13)~~~0.885 (13) & 0.870 (12)~~~0.880 (12) & 0.868 (14)~~~0.881 (14) \\
                   &                       &                       &                       \\
\hline
\end{tabular}
\end{center}

$^a$ only S1, S2, B1, B2 data sets; \hspace{0.4cm} $^b$  only S1, S2, R, B1, B2 data sets;\\
$^c$ errors in data sets R, B1, B2, M enlarged by a factor of 2
\end{table}

\end{document}